\documentstyle[preprint,tighten,eqsecnum,epsfig,aps]{revtex}

\def\beq{\begin{equation}}   \def\eeq{
\end{equation}}

\begin{document}
\title{On the limiting behaviour of  hard processes in QCD at small $x$.
\thanks{Talk given by B.Blok at second conference: QCD at 
cosmic energies, Skopelos, Greece, September 2005. }}
\author{B.Blok\thanks{E-mail:blok@physics.technion.ac.il}\address{Physics
Department, Israel Institute of Technology, 32000 Haifa, Israel}}
\author{ L. Frankfurt\thanks{E-mail: frankfur@lev.tau.ac.il} }
\address{School of Physics and Astronomy, Raymond and Beverly Sackler
Faculty of Exact Sciences, Tel Aviv University, 69978 Tel Aviv,
Israel} \maketitle \thispagestyle{empty}

\begin{abstract}

At sufficiently small $x$ where pQCD methods become insufficient we rewrote sum of
dominant Feynman diagrams for single-scale hard processes in the form of the
effective field theory where a quasiparticle corresponds to pQCD ladder. We explain
that  pQCD series are divergent at small $x$ since pQCD produces tachyon in
effective field theory and because of degeneracy of vacua . Within the WKB
approximation we found the
hierarchy among quasiparticle interactions due to smallness of the QCD coupling
constant which justifies restriction by the triple quasiparticle interaction in
the Lagrangean of effective field theory. Degeneracy of vacua is removed by accounting
for the classical solutions -kinks which cannot be decomposed into series over powers
of $\alpha_s$ but describe color inflation, Bose-Einstein condensation of
quasiparticles.  Quantum fluctuations around the WKB solution lead to the spontaneous
violation of two-dimensional translation symmetry and the appearance of zero modes-
``phonons'' which are relevant for the black disc behaviour of small x processes.
Account of gluon exchanges between overlapping ladders neglected in the first
approximation produces a color network occupying a "macroscopic" longitudinal volume. We
discuss briefly possible role of new QCD phenomena in the two-scale hard processes .

 \end{abstract}

\setcounter{page}{1}
\section{Introduction}

\par One of the challenging properties of the perturbative QCD is the conflict between
predicted within DGLAP approximation rapid increase of cross
sections of hard processes with energy \cite{Gross,Dokshitzer} and
the probability conservation. The QCD evolution equation describes
well the experimental data ( see review \cite{AC} where
$F_2p(x,Q^2) \propto x^{-\mu(Q^2)}$ for $\mu(Q^2) \approx 0.2 $
and $Q^2\approx 10 GeV^2$). The slope $\mu$ is increasing with
$Q^2$ . Cross sections of diffractive processes like diffractive
electroproduction of vector mesons and photoproduction of hidden
heavy flavor mesons increase even faster $\propto x^{-2\mu(Q^2)}$
\cite{BGFMS94,CFS} . This behaviour has been observed recently
(for the review and references see ref. \cite{AC}). Predicted by
DGLAP increase of parton distributions with energy is too fast to
keep evident condition $\sigma(tot)\ge \sigma(diff)$ for cross
sections of hard processes i.e. the probability conservation
\cite{AFS}. The pQCD factorization theorems contradict the
Froissart type bound for the cross-section of scattering of
spatially small colorless dipole that follows from the nonlinear
relations between the Green functions in QCD and analytic
properties of amplitudes in the plane of momentum transfer
\cite{MFS}: \beq \sigma (s)\sim \log^2(s/s_o) \label{fr} \eeq
where s is the center of mass energy of colliding particles. The
same puzzle exists in the other approximations such as LO and NLO
BFKL approximation \cite{BFKL,Ciafaloni}.  The closely connected
property of pQCD is the fast increase of the correlators of local
currents with distance \cite{BF1,BF2} . The presence of long range
interaction suggests the instability of the pQCD vacuum. The
puzzles reveal itself in the hard phenomena where the virtualities
of the interacting partons are large and therefore they are
unrelated to the poorly understood physics of quark confinement
and spontaneous broken chiral symmetry.

\par
The pQCD calculation within the DGLAP approximation
\cite{Strikman} of the scattering of the color-neutral two-gluon
dipole of size $\approx 1/Q$ off a proton target violates  the LT
approximation at $x\le x_{\rm cr}(Q^2)\sim 10^{-5}$--$10^{-4},$
i.e. in the kinematics conveniently achievable at LHC. (The estimate is
given for the scale of hard processes $Q^2\sim 10$~GeV$^2$ chosen
to guarantee the smallness of the running invariant charge.) At
these scales the higher twist contributions blow up. This scale of
$x$ seems to be significantly larger than the values of $x$ where
different pQCD approximations start to diverge in the predicted
behavior of the amplitudes of the hard processes \cite{Salam}.
Thus the violation of the LT approximation ,of DGLAP,LO+NLO BFKL..
approximations , and maximal value permitted by probability
conservation for the amplitudes of hard processes are achieved  at
the values of $x$ where pQCD calculation of LT contributions is
still more or less unambiguous. Thus the challenging problem is to
develop technology of calculations and to predict existence and
probability of new QCD hard phenomena at small $x$ where
conventional pQCD approaches should fail.

\par
To simplify this difficult problem we restrict ourselves by the
theoretical analysis of the one-scale hard processes like
$\gamma^*(Q^2)+\gamma^*(Q^2)\rightarrow $ hadrons, where $Q^2$
evolution is suppressed. Account of the running of the coupling
constant and, to some extent, energy-momentum conservation
suppress fast diffusion to large and small transverse distances
\cite{Ciafaloni}. In our consideration we will neglect diffusion
to small distances since in any case it can not change properties
of amplitudes near black body limit. However we will take into
account V.Gribov diffusion to large impact parameters which is
distinctive feature of ladder diagrams \cite{Gribovdiff} .
Moreover we restrict ourselves by the consideration of the
kinematics of extremely small $x$   where diffractive production
of large mass states is not hampered by energy-momentum
conservation. In this kinematics assumption on the dominance of
shadowing due to diffraction into small masses  will be hardly
consistent with the probability and energy-momentum conservation
cf.\cite{FSW}. So we assume dominance of processes of diffraction
into large invariant masses.

\par
The aim of this talk is to calculate limiting behaviour of small x
processes. Following V.Gribov Reggeon Calculus
\cite{GribovCalc,VGribov-AMigdal} we develop  effective theory
(EFT) describing the single scale hard processes and solve it
within the WKB approximation. This approach helps to visualize the
challenging problem: modern methods of pQCD produce tachyon in EFT
and as in the theory of superconductivity \cite{LL} and in the
string model in 26 dimensions \cite{Shatashvily} it is necessary
to find the physical vacuum state of EFT and then to develop the
perturbation theory. The ideas and methods of spontaneously broken
continuous symmetries will be heavily used also.  The new QCD
phenomena we find may appear important for many-scale hard
processes as well even if methods developed in the paper are not
directly applicable.

\par 
We work in the kinematics where light-cone wave functions of the colliding $\gamma^*$ are
dominated by configurations having large invariant mass and containing many constituents with
large transverse momenta  ---the hard-QCD analogue of the triple-reggeon  limit. Thus a necessary
 kinematical condition of applicability of our method is $x\le x_{\rm cr}(Q^2)\cdot 10^{-2}$,
i.e. $x\sim 10^{-6}-10^{-7}$. Here $x_{\rm cr}(Q^2)$ can be determined from the condition  that
the contribution of one ladder in dipole--dipole scattering at central impact parameters reaches
the black disk limit. The demand for the existence of the three-reggeon limit is  the main
kinematic limitation for the applicability of our approach.

\par 
To account for the coherence of high-energy processes and  the rapid 
increase with energy of
the amplitudes of hard small-$x$ processes, we construct an
effective field theory (EFT) of interacting perturbative ladders,
which are our quasiparticles and neglect initially by gluon
interactions between ladders . We take account of the interaction
between quasiparticles along the lines of Gribov's Reggeon field
theory \cite{GribovCalc}. (This approach is in the spirit of
statistical models of critical phenomena which account for the
interactions between major modes only. Specifics of the physics of
large longitudinal distances are included in the concept of
quasiparticle.) The interaction between quasiparticles when the
amplitudes of hard processes are near the unitarity limit can be
easily evaluated when the WKB approximation is combined with the
smallness of the running coupling constant in pQCD . We showed
that smallness of the running coupling constant helps to establish
a hierarchy of multiladder interactions and the dominance of the
triple-ladder vertex \cite{BF3}. The smallness of the multiladder
vertices in pQCD implies that basic phenomena characteristic of
the BDL should be insensitive to the restriction by the
triple-ladder vertex. This observation helps to fix the form of
the Lagrangian of the EFT. Moreover smallness of effective triple
ladder vertex justifies applicability of semiclassical
approximation. Thus EFT is solvable within WKB approximation and
leads to transition to BDL- "black disk limit". Derived in our
papers Lagrangean of EFT is almost identical to that analysed in
preQCD Reggeon Calculus \cite{amati1,amati2,amati3,amati4} long
ago. So in many situations we may translate results obtained in
these papers to  the language appropriate for the theoretical
description of  single scale hard QCD regime.

\par 
It follows from EFT that the transition to BDL in the one scale hard processes 
is  a result of
the existence of nonperturbative solutions of equations of EFT -
kinks in rapidity-impact parameter space. Transition due to the
general EFT kink is suppressed by the factor $\sim
\exp(-1/N_c\alpha_s)$. The critical kinks that correspond to
actual transitions between vacua are the limiting cases of the
families of these noncritical kinks.

\par The transition to BDL is of the inflationary type:  in the case of collisions of two small
dipoles  the time scale $T_I\sim \exp(1/\mu(Q^2))/Q$ of the transition significantly  smaller than the
time needed for the formation of perturbative ladder $\propto  1/Qx^{1-\mu(Q^2)}$.

\par  The nonperturbative transition produces ladders which strongly overlap in the impact parameter
space.  Due to exchange by constituents between overlapping ladders system of ladders becomes color
network .\cite{BF3} However to evaluate properties of this color network, it is necessary to describe
the BDL transitions (the  kinks) directly in terms of the QCD language, which is yet not done.

\par To summarise, we were able to show that QCD can be formulated as solvable in quasiclassical
approximation effective EFT (analogue of V.Ginzburg- L.Landau theory of superconductivity) , and
this EFT has a transition to the BDL limit, which in the QCD language is a color network.

\par  For two scale processes like small dipole scattering off hadron(nucleus) target related effects
should reveal itself at extreemly large energies where QCD evolution is restricted by dipole
fragmentation region. The challenging question is to find quasiparticles  which dominated at lesser
energies.

\section{Effective field theory in QCD}

\par   We assume, as it was mentioned above, that the dominant degrees of freedom--the
quasiparticles of EFT, are pQCD color-singlet ladders -"Pomerons" \cite{BF3}.
In the derivation of the EFT from Feynman diagrams we follow Gribov's Reggeon calculus
\cite{VGribov-AMigdal,GribovCalc} and take advantage of the simplifications due
to the smallness of the running coupling constant.

The equations of the EFT  can be derived  as Lagrange equations of
motion from an effective Lagrangian that has a form
\begin{eqnarray}
L&=&p\partial_y q -q\partial_y p -\alpha'p\triangle_b q- \mu
pq-\kappa pq(p+q) - c_{\rm dipole}\int \exp(-BQ/2)
 q(y,b-B)d^2 B \delta(y)\nonumber\\[10pt]
&-&c_{\rm dipole}\int \exp(-BQ/2) p(y,b-B)d^2B \delta(y -Y)
\nonumber\\[10pt]
\label{lag}
\end{eqnarray}
Here  $c_{\rm dipole}$ accounts for the normalization of the virtual photon
wave function.  The first three terms have a straightforward interpretation
in the case of noninteracting  quasiparticles. They follow from the Mellin
transformation of the Green function of the free quasiparticle,
$G=[j-1-\mu(Q^2)-\alpha' k^2]^{-1}$, in the plane of complex angular momentum
$j$ in the crossed channel. Since$\mu\ge 0$ Green function has pole in the
unphysical region forbidden by probability conservation and analytic
properties of amplitudes in the plane of momentum transfer t i.e. pQCD "Pomeron"
is tachyon of EFT. The fields $p(y,b)=\psi^+$ and $q(y,b)=\psi$ are the quasiparticle
fields, analogous to Gribov's Pomeron fields. We denote
$\partial_y=\partial_{\log(x_{0}/x)}$ where $y$ is rapidity and
$x=Q^2/(2pq)$.  The quantity $x_0\approx 0.1-10^{-2}$ denotes the length
of the fragmentation region where there are no $\log(x_0/x)$
factors. $ F_{2p}(x,Q^2),\ xG_{p}(x,Q^2)\sim (x_0/x)^{\mu
(Q^2)}$ with $\mu>0$; $\mu(Q^2)\sim\alpha_sN_c/\pi+...$
has been calculated in pQCD (for a review and references see Ref.~\cite{Salam}).

\par 
The third term describes the dependence on  the collision
energy of the essential  impact parameters.  We assume that it has
the form  $\alpha'_{P}p\triangle_bq$ natural for ladder diagrams
\cite{GribovCalc}. Here $\vec b$ is a two-dimensional impact
parameter,and $\alpha_{P}'$ is the 'Pomeron' slope . $\alpha_{P}'$
is  small within pQCD- $\alpha_{P}'\propto N_c
\alpha_s(Q^2)/Q^{2}$. At the same time, near the BDL the effective
$\alpha_{P}'$ cannot be small \cite{Strikman}. Exact form of this
term is unknown since it is sensitive near unitarity limit to
unknown nonleading order terms in running coupling constant . So
we choose this term to account for V.Gribov diffusion within pQCD ladder.

\par 
The evaluation within the WKB approximation of multi-Pomeron vertices near the BDL
\cite{BF3} shows that,  the relative contribution of the fourth and higher multi-Pomeron
vertices is suppressed by powers of $\alpha_s$ compared to the triple-ladder term. Thus
for the description of hard QCD phenomena it should be sufficient to
restrict ourselves to the triple-ladder interaction. In the lowest
order in the coupling constant, the triple-Pomeron vertex,
 \beq
L_4=\kappa pq(p+q),\,\,\,\kappa \propto  i\frac{N_c^2
\alpha_s^2}{\lambda}
\label{11}
\eeq
is due to the interaction of ladders via one gluon loop. This estimate,
accounts QCD evolution, a running coupling constant, and Sudakov form factors
which suppress nonperturbative contribution discussed in  Ref.~\cite{Bartels},
The existence (but not the properties) of new QCD phenomena is not sensitive
to the actual  value of $\lambda \approx Q$ which is the characteristic transverse
momentum of the constituents of the pQCD ladder where it splits into two new ladders.

\par
We neglect eikonal-type inelastic rescatterings since a bare
particle may have one inelastic collision and any number of
elastic collisions only, \cite{Mandelstam,GribovCalc}. For the
interactions that rapidly increase with energy, requirements of
causality, positivity of probability for physical processes, and
energy-momentum conservation can be hardly satisfied within such a
set of diagrams \cite{FSW}. In contrast, the contribution of
rescatterings  due to an inelastic diffraction into mass $M^2$,
where $\beta=Q^2/(Q^2+M^2)$, is not too small, dominates in
two-scale hard small-$x$ phenomena at $x\approx x_{cr}$. We
include this contribution in the scale factor of the source.

\par  
EFT is formally different from the approaches suggested in Refs.~\cite{A.Mueller}
\cite{McLerran}, \cite{Iancu}. EFT accounts for the increase with energy of essential impact parameters,
neglect elastic eikonal rescatterings etc. However major difference is in the account of kinks and
quantum fluctuations around kinks, of  the phenomenon of spontaneously broken
continuous symmetry.  Moreover within EFT  the contribution of the quantum fluctuations around
the pQCD vacuum is negligible in one-scale hard processes.

\par 
Within the approximations made in this paper the coupling
of the pQCD ladder to a hadron can be treated as  the interaction
with a source. The actual form of these term
is unimportant for most of the results obtained in this paper, so we
refer a reader to ref. \cite{BF3} for detailes and here just
include this interaction into Lagrangian \ref{lag}.

\section{Critical phenomena in hard QCD near unitarity limit}

\par The form of Lagrangian (\ref{lag}) relevant for hard QCD phenomena coincides
with the  particular preQCD model for the Lagrangian of V.Gribov Reggeon Field Theory
analysed in refs. \cite{amati1,amati2,amati3} . Thus in the further analysis we
use WKB  solution of Lagrangian equations of motion and quantum fluctuations around them
found in refs. \cite{amati1,amati2,amati3,amati4,amati5} but interpret them in
terms of hard QCD phenomena .

\par Let us formulate here the main properties of the WKB solution .

\par Distinctive property  of the Lagrangian (\ref{lag}) is the
existence, apart of the usual perturbative vacua
\beq
p=0,q=0,
\label{v0}
\eeq
 , of the two new vacua:
 \beq
p=\mu /\kappa, q=0\,\,\,\,{\rm and}\,\,\,\,\, p=0, q=\mu /\kappa .
\label{v1} \eeq

\par  
The detailed analytical analysis of the model is possible in 1+1 dimensions
only, where the equations of motion  are reduced to ordinary differential equation.
In refs. \cite{amati1,amati2,amati3} the family of kinks, characterized by a 2d velocity
parameter v has been found. These kinks interpolate between 3 vacua  eqs. (\ref{v0},\ref{v1}).
The action of the kink is finite $S\sim (\mu/\kappa)^2(2-v)\phi_0^2$, where $\phi_0$ is the
field value at $b=vY$ and $v $  is the kink velocity.  It  is proportional to
$1/(N_c^2\alpha_s^2)$, where we used the dependence of $\mu$ and $\kappa$ on $N_c$ discussed
in section 2. For the value of parameter $v=2$ we obtain critical kinks with zero action. The
existence of these critical kinks is crucial for the quantisation of the theory. The classical
contributions of these kinks into wave function are not exponentially suppressed.
Quantum fluctuations around these kinks are described by a positive quadratic form, cf. \cite{amati3}.

\par 
The characteristic property of  kinks is their step function form.
One of the functions p or q behaves approximately like a step function
\beq
p (q)\sim \theta (v\sqrt{\alpha'\mu} (y-y_0)-\vert \vec b-\vec b_0\vert)
\label{cf}
\eeq
where v is the kink velocity (a free parameter, v=2 for critical kink).
The solution contains arbitrary parameter $y_0$ that helps
to understand why the physics related to the fragmentation
can be hidden into the properties of the source.
The arbitrary solution depends also on $\vec b-\vec b_0$. The value
of $b_0$ is not fixed by equations. This is a zero mode relevant for
the appearance of "phonons" in quantum fluctuations.

\par  
In physical 2+1 dimensions there is no analytical expression for the
solution giving critical
kinks with zero action as well as a full classification of kinks .
So in 2+1 dimensions we rely on the results of the numerical
simulations  made in \cite{amati1,amati2},  which found the same
properties of kinks for the 2+1 dimensional theory as for 1+1
dimensional one.

\par 
The knowledge of the family of kink solutions permits
semiclassical quantization of the theory and to calculate  S-matrix.
\par Remarkable property of the quantum fluctuations around
 critical kinks is the existence  of zero modes-"phonons" in EFT
 , which are characterized by the linear dispersion :
\beq
E=i2\sqrt{\alpha'\mu}k=2\sqrt{\alpha'\mu} P_{\rm cl}
\label{C1}
\eeq
where $P_{\rm cl}=\int d^2b p\frac{dq}{db}$ is the total classical
momentum derived from the EFT action.
  (Energy and momentum of kinks  are defined as components of $\int d^2b T^{00}$
and $\int d^2b T^{0i}$, where  the energy momentum tensor of
the EFT is found via the Nether theorem).

\par 
In addition, there exists a band of low lying states with
a dispersion relation $E\sim k^2$, and a band of states with a gap
$\sim \mu$. It can be argued that the first set of states
corresponds to unshifted quasiparticle fluctuations around
perturbative vacuum (\ref{v0}) in the presence of a kink and viewed
from the reference frame moving with a critical speed $v_0$. The
higher modes can be interpreted as collective fluctuations of a
condensate of ladders, i.e. quasiparticles interacting with kinks.
It is easy to prove that these modes do not influence the expanding
disk solution asymptotically, although can be important outside
the disk, and near the transition to a black disk regime.  Only solutions
with linear spectrum are relevant for the asymptotic behaviour
of high energy processes.

\par 
These results are valid in 1+1 dimensional model that was solved analytically and are
confirmed by the numerical analysis of spectrum of quantum fluctuations around kinks that  has been
performed  in physical 2+1 dimensional case in refs. \cite{amati1,amati2}.  Moreover, it
was proved on the discrete lattice that this model can be continuously connected with the Ising
model in transverse magnetic field.

\par 
To summarize, the quasiclassical solution of EFT has following
distinctive features (see detailes in ref. \cite{BF3}):

\par 
A) EFT has three  degenerated "vacua" (\ref{v0}),
 (\ref{v1}), (The word "vacua" is in the brackets, since EFT is the
theory with nonhermitian Hamiltonian).
The true wave function of the physical basic state
is a linear combination of these three vacua,  and the Hamiltonian
is  diagonalized by "critical" kinks with zero action. If the initial
state is a perturbative one interacting with a source,
 it evolves in rapidity and becomes a
 condensate of quasiparticles=ladders.
\par B) The S-matrix is given by the functional integral
$ S(B,Y)=\int dp\int dq \exp(-L)$ where
the corresponding action is calculated via the classical kink
solution described above, but includes now the source terms-the
coupling with virtual photons. The main contribution to S-matrix
comes from quantum fluctuations around critical kinks with zero
action.  Spectrum of excitations begins from massless excitations-
"phonons". The contribution of the kinks into the two-
quasiparticle Green functions that determines the partial waves
is: \beq D(B,Y)=(\mu/\kappa)^2\theta(B^2-\alpha'\mu Y^2).
\label{v12} \eeq  . $(\kappa/mu)^2 D$ as given by eq. (\ref{v12}) is 1
inside the black disk and  suppressed exponentially as
$\mu/\kappa \sim \sim 1/\alpha_s$, outside.

\par 
C) The obtained state is described by the asymptotic wave function
\beq
\Psi(y)=\exp(-\frac{\mu}{\kappa}\int d^2bq(b,y)
\theta(b^2-\alpha'\mu Y^2)\vert \phi_0>
\label{slon3}
\eeq
Here $\vert\phi_0>$ is a perturbative vacuum. To the extent that
correlations may be neglected the asymptotic vector (\ref{slon3})
is a coherent state \cite{amati3} and the S-matrix is given by
\beq
S(B,Y)=\exp(-\frac{\mu}{\kappa}\theta(2\sqrt{\alpha'\mu}Y-B)),
\label{slon31}
\eeq
where we explore that the target is localized near the impact parameter
$b\sim B$. This is just the Froissart (BDL) behaviour.
  We want to draw attention that $\Psi(y)$ can not be
obtained by decomposition over powers of $\alpha_s$.
\par D)  In other words in the limit of infinite energies
$Y\rightarrow \infty$  produced state corresponds to a Bose-Einshtein
condensate of  ladders  in the entire space. However,for finite
energies the solution is the black disk of radius
$R^2\sim R_0^2+\alpha'\mu Y^2$. The equation (\ref{slon3}) gives the
exact form of the wave function of the Bose-Einstein condensate of
ladders as a function of rapidity Y.

\par E) The transition occurs for given impact parameter b at
rapidities $b^2\sim \mu Y(1/\lambda^2+4\alpha'_PY)$.

\section{Kinks and QCD.}

Some properties of classical solutions of EFT can be
 understood directly in QCD.  A kink produces action proportional to
$(\mu/\kappa )\sim (1 /N_c\alpha_s)$ in some power.
  The dependence  of the S-matrix (\ref{slon31}),
 of the critical kink action (understood as the limit of a family of
 kinks with nonzero action)  on the coupling constant $\alpha_s$,
  the spontaneously broken translational invariance,
 and existence of "phonon" show that this
is a novel nonperturbative QCD phenomenon.
\par The characteristic form of a kink is the step-function in Y
space, with the width
of the order $\delta Y\approx \log (\delta E/Q) =1/\mu$. Thus
coherent length relevant for the evolution of kink is enhanced
due to large Lorentz slowing down interaction factor as
$T_l \sim \delta E/Q^2\approx exp (1/\mu)/Q\approx 10^2/Q$.
In other words, for sufficiently low x $T_I\ll T_c$, where $T_c$ is coherence
length $T_c\sim 1/(Q x^{1-\mu})$.
(This formulae differs from more familiar LT formulae  $T_c\sim 1/(2m_N x)$. It
follows from the violation of LT approximation in the kinematics near BDL.)
This rapid transition  to BDL can be called the "color inflation": one ladder
due to the tunneling transition blows up during time $T_I$ and
creates an entire region of space filled with gluon ladders.
During time $T_l$   $\sim (\mu^2/\kappa^2)R(Y)^2$
ladders are created , where $R(Y)$ is a black disk radius for
a given rapidity $Y$.
\par It is easy to evaluate density of ladders in coordinate
space by solving discussed above diffusion equations cf. similar
analysis in \cite{amati3} One obtains
\beq
d_t^2 \sim N_c
\alpha_s/\lambda ^2
\label{v**}
\eeq
Thus $ l_t^2/d_t^2\approx \alpha_s N_c \ll 1 $,where
$l_t\sim \kappa/\mu\sim \alpha_sN_c/\lambda$
is the characteristic scale of a ladder in a transverse parameter
space. It follows from above
estimate that pQCD ladders overlap significantly .  But
overlapping ladders can exchange by quarks and gluons  because
pQCD  does not produce barriers between ladders. This distinctive
feature of color network macroscopical in the longitudinal
direction resembles quark-gluon plasma .

\par  Understanding  the actual longitudinal structure of the
system can not be resolved within EFT. Indeed, even after the phase
transition, the ladders continue to grow till color network
achieves longitudinal length $1/(Qx^{1-\mu})$ .

\section{Observable phenomena.}

\par  Investigation of one hard scale processes will be subject of research
at the next generation of e$\bar e$ colliders. Two scale processes like
structure functions of hadron (nuclear) target are dominated
by $Q^2$ evolution at significant region of small $x$.  At extreemly small
$x\ll x_{cr}10^{-2}$  $Q^2$ evolution will be restricted by the
virtual photon fragmentation region . In this kinematics
which can be achieved in cosmic ray physics
QCD phenomena found in the paper may reveal itself.
This contribution is additional  to the contribution of large and
moderate masses $M^2$ in the photon w.f. where
$\beta=Q^2/(Q^2+M^2)$ is not too small.
Color inflation may reveal itself as threshold like
increase of multiplicity of hadrons and softening of hadron
distribution in longitudinal direction. Hard QCD phenomena discussed
in the paper may reveal itself  in the central pp,pA and AA collisions
in the regime where hard collisions are near the BDL.

Promising way to identify the onset of the new QCD regime
would be to measure Mueller-Navelet process \cite{MN}: $p+p\to
jet+X+jet$ where the distance in the rapidity between high $p_t$
jets is large. Expected behaviour is:  initial fast increase of cross
section with $y$ predicted by pQCD should change to the fast
decrease at larger $y$ because of the color inflation, i.e.
disappearance of the long range correlations in the rapidity
(coordinate) space near the BDL.

In the case of central heavy ion collisions one may put $\alpha_{P}'=0$
because of large radius of heavy nuclei. So color inflation will reveal
itself as the threshold behaviour in the hadron production,
leading to the formation of the color network at sufficiently large energies .
\par {\bf Acknowledgements.} We thank Prof. M. Strikman for useful
discussions. The materials of the conference are located at the
WEB site http://www.lpthe.jussiue.fr/Skopelos.

\end{document}